%% file: Lec_Notes_final.tex
\documentclass[journal]{IEEEtran}

\input{myshorts}

\usepackage{amsfonts,amsthm,amsmath,amssymb}
\usepackage{graphicx,subcaption}
\usepackage[noadjust]{cite}
\usepackage[ruled,vlined]{algorithm2e}
\usepackage[dvipsnames]{xcolor} 
\usepackage[bookmarks,colorlinks]{hyperref}
\usepackage{acronym}
\DeclareMathAlphabet{\pazocal}{OMS}{zplm}{m}{n}
\DeclareMathOperator*{\argmin}{argmin}
\DeclareMathOperator*{\argmax}{argmax}

\input{Preamble.tex}

\begin{document}
\title{Discriminative and Generative Learning for Linear Estimation of Random Signals [Lecture Notes]}
\author{Nir Shlezinger and
 Tirza Routtenberg
 \thanks{
		The authors are with the School of ECE, Ben-Gurion University of the Negev, Beer-Sheva, Israel (email: \{nirshl, tirzar\}@bgu.ac.il).
}
 }

	\maketitle
\graphicspath{{figures/}}

\section{Scope}
Inference tasks in signal processing are often characterized by the availability of reliable statistical modeling with some missing instance-specific parameters. One conventional approach uses data to estimate these missing parameters and then infers based on the estimated model. Alternatively, data can also be leveraged to directly learn the inference mapping end-to-end. These approaches for combining partially-known statistical models and data in inference are related to the notions of generative and discriminative models used in the machine learning literature~\cite{ng2001discriminative, jebara2012machine}, typically considered in the context of classifiers.  

The goal of this lecture note is to introduce the concepts of generative and discriminative learning  for inference  with a partially-known statistical model. While machine learning systems often lack the interpretability of traditional signal processing methods, we focus on a simple setting where one can  interpret and compare the approaches in a tractable manner that is accessible and relevant to signal processing readers. 
In particular, we exemplify the approaches for the task of Bayesian signal estimation in a jointly Gaussian setting with the \ac{mse} objective, i.e., a linear estimation setting. Here, the discriminative end-to-end approach directly learns the linear minimum \ac{mse} (LMMSE) estimator, while the generative strategy yields a two-stage estimator, which first uses data to fit the linear model, and then formulates the LMMSE estimator for the fitted model. The ability to derive these estimators in closed-form facilitates their analytical comparison. It is rigorously shown that  discriminative learning results in an estimate which is more robust to mismatches in the mathematical description of the setup. Generative learning, which utilizes prior knowledge on the distribution of the signals, can exploit this prior to achieve improved \ac{mse} in some settings. These analytical findings are numerically demonstrated in a numerical study, which  is available online as a Python Notebook, such that it can be presented alongside the lecture detailed in this note.




\section{Relevance}
Signal processing algorithms traditionally rely on mathematical models for describing the problem at hand. These models correspond to domain knowledge obtained from, e.g., established statistical models and understanding of the underlying physics. In practice, statistical models often include parameters that are unknown in advance, such as noise levels and channel coefficients, and are estimated from data.

Recent years have witnessed a dramatic success of machine learning, and particularly of deep learning, in domains such as computer vision and natural language processing \cite{lecun2015deep}. For inference tasks, these data-driven methods typically learn  the inference rule directly from data, rather than estimating missing parameters in the underlying model, and can  operate without any mathematical modeling.  Nonetheless, when one has access to some level of domain knowledge, it can be harnessed to design inference rules that benefit over black-box approaches in terms of performance, interpretability, robustness, complexity, and flexibility \cite{shlezinger2022model}. This is achieved by formulating the suitable inference rule given full domain knowledge, and then using data to optimize the resulting solver directly, with various methodologies including learned optimization \cite{agrawal2021learning}, deep unfolding \cite{monga2021algorithm},  and the augmentation of classic algorithms with trainable modules \cite{shlezinger2021model}.

The fact that signal processing tasks are often carried out based on partial domain knowledge, i.e., statistical models with some missing parameters, and data, motivates inspecting which design approach is preferable: the model-oriented approach of using the data to estimate the missing parameters, or the task-oriented strategy, which leverages data to directly optimize a suitable solver in an end-to-end manner? These approaches can  be related to the notions of {\em generative learning} and {\em discriminative learning}, typically considered in the machine learning literature in the context of classification tasks \cite[Ch. 3]{theodoridis2020machine}. In these lecture notes, we address the above fundamental question for an analytically tractable setting of linear Bayesian estimation, for which the approaches can be rigorously compared,  connecting machine learning concepts with interpretable signal processing practices and techniques.


\section{Prerequisites}
This lecture note is intended to be as self-contained as possible and suitable for the undergraduate level without a deep background in estimation theory and machine learning.  As such, it requires only  basic knowledge in probability and calculus. 


\section{Problem Statement} 
\label{sec:Problem}
To formulate the considered problem, we first review some basic concepts in statistical inference, following~\cite{shalev2014understanding}. Then, we elaborate on model-based and data-driven approaches for inference. Finally, we present the running example considered in the remainder of this lecture note of linear estimation in partially-known measurement models.

\subsection{Statistical Inference}
 The term {\em inference} refers to the ability to conclude based on evidence and reasoning.
    While this generic definition can refer to a broad range of tasks, we focus in our description on systems that estimate or make predictions based on a set of observed measurements.  In this wide family of problems, the system is required to   map an input variable $\Input$, taking values in an {\em input space} $\InputSpace$  into a prediction of a target variable $\Label$, which takes value in the {\em target space} $\LabelSpace$. 
    
     The inputs are related to the targets via some statistical probability  measure, $\Distribution$, referred to as the {\em data  generating} distribution, which is defined over $\InputSpace \times \LabelSpace$.  
    Formally, $\Distribution$ is a joint distribution over the domain of inputs and targets. One can view such a distribution as being composed of two parts: a distribution over the unlabeled input $\Distribution_\Input$, which sometimes is called the {\em marginal distribution}, and the conditional distribution over the targets given the inputs $\Distribution_{\Label|\Input}$,  also referred to as the {\em discriminative} or {\em inverse} distribution.

{\bf Inference rules}  can thus be expressed as mappings of the form
    \begin{equation}
        \label{eqn:mapping}
        f:\InputSpace \mapsto  \LabelSpace.
    \end{equation}
    We write the decision variable for a given input $\Input\in\InputSpace$ as $\hat{\Label} = f(\Input)\in\LabelSpace$. The space of all possible inference mappings of the form \eqref{eqn:mapping} is denoted by $\mySet{F}$.
    The fidelity of an inference mapping is measured using a loss function
    \begin{equation}
        \label{eqn:lossfunction}
        l:\mySet{F}\times\InputSpace\times\LabelSpace \mapsto \mathbb{R},
    \end{equation}
    with $\mathbb{R}$ being the set of real numbers.
    We are generally interested in carrying out inference that minimizes the {\em risk function}, also known as the {\em generalization error}, given by:
    \begin{equation}
        \label{eqn:risk}
        \mySet{L}_{\Distribution}(f) \triangleq \E_{(\Input,\Label)\sim \Distribution} \{l(f,\Input,\Label)\},
    \end{equation}
    where $\E\{\cdot\}$ is the stochastic expectation. 
     Thus, the goal is to design the inference rule $f(\cdot)$ to minimize the generalization error $\mySet{L}_{\Distribution}(f)$ for a given problem.  

\subsection{Model-Based versus Data-Driven}
\label{subsec:MBvsDL}
The risk function in \eqref{eqn:risk} allows to evaluate inference rules and to formulate the desired mapping as the one that minimizes $\mySet{L}_{\Distribution}(f)$.
The main question is how to find this mapping. \textcolor{NewColor}{Approaches to design $f(\cdot)$ can be} divided into two main strategies: the statistical model-based strategy, referred to henceforth as {\em model-based}; and the pure machine learning approach, which  relies on data, and is thus referred to as {\em data-driven}. The main difference between these strategies is what information is utilized to tune $f(\cdot)$. 

{\bf Model-based methods}, also referred to as {\em hand-designed schemes}, set their inference rule, i.e.\ tune $f(\cdot)$ in \eqref{eqn:mapping}, to minimize the risk function $\mySet{L}_{\Distribution}(\cdot)$, based on full domain knowledge.  The term {\em domain knowledge} typically refers to  prior knowledge of the underlying statistics relating the input $\Input$ and the target $\Label$, where {\em full domain knowledge} implies that the joint distribution $\Distribution$ is known. For instance, under the  squared-error loss, $ l_{\rm MSE}(f,\Input,\Label) = \|\Label - f(\Input)\|_2^2$, the optimal inference rule is the {\em  minimum \ac{mse}} estimator, given by the conditional expected value $f_{\rm MMSE}(\Input) = \E_{\Distribution_{\Label|\Input}}\{\Label | \Input\}$, whose computation requires knowledge of the discriminative distribution   $\Distribution_{\Label|\Input}$. 

 Model-based methods are the foundations of statistical signal processing. However,  in practice, accurate knowledge of the distribution that relates the observations and the desired information is often unavailable. Thus, applying such techniques commonly requires imposing some assumptions on the underlying statistics, which  in some cases  reflects the actual behavior, but may also constitute a crude approximation of the true dynamics. Moreover, in the presence of inaccurate model knowledge,   either as a result of estimation errors or due to enforcing a model which does not fully capture the environment, the performance of model-based techniques tends to degrade. Finally, solving the coupled problem of  model selection and estimation of the parameters of the selected model is a difficult task, with non-trivial inference rules  \cite{Harel_Routtenberg_Bayesian,Meir_Routtenberg}.

{\bf Data-driven methods} learn their mapping from data rather than from statistical modeling. This approach builds upon the fact that while in many applications coming up with accurate and tractable statistical modeling is difficult, we are often given access to data describing the setup. In a supervised setting considered henceforth, data is comprised of a training set consisting of $\Ntraining$ pairs of inputs and their corresponding target values, denoted by $\Data = \{\Input_t, \Label_t\}_{t=1}^{\Ntraining}$. 

Machine learning provides various data-driven methods which form an inference rule $f$ from the data $\Data$. Broadly speaking and following the terminology of \cite{ng2001discriminative}, these approaches can be divided into two main classes:
\begin{enumerate}
    \item {\em Generative models} that use $\Data$ to estimate the data generating distribution $\Distribution$. Given the estimated distribution, denoted by $\hat{\Distribution}_{\Data}$, one then seeks the inference rule which minimizes the risk function with respect to $\hat{\Distribution}_{\Data}$, i.e., 
    \begin{equation}
        \label{eqn:MBOpt}
        f^* = \argmin_{f\in \mySet{F}} \mySet{L}_{\hat{\Distribution}_{\Data}}(f),
    \end{equation}
    where $\mySet{L}_{\Distribution}(f)$ is defined in \eqref{eqn:risk}.
    \item {\em Discriminative models} where data is used to directly form the inference rule as a form of {\em end-to-end learning}. Without access to the true distribution $\Distribution$, one cannot directly optimize the risk function \eqref{eqn:risk}, typically resorting to the {\em empirical risk} given by 
   \begin{equation}
   \label{eqn:EmpRisk}
       \mySet{L}_{\Data}(f) \triangleq \frac{1}{\Ntraining}\sum_{t=1}^{\Ntraining} l(f,\Input_t, \Label_t).
   \end{equation}
   
   To avoid overfitting, i.e., coming up with an inference rule which minimizes \eqref{eqn:EmpRisk} by memorizing the data, one has to constrain the set $\mySet{F}$. This requires  imposing a structure on the mapping, which is often dictated by a set of parameters denoted by $\myVec{\theta}$, taking values in some parameter set $\Theta$, as considered henceforth.  Thus, 
   the system mapping is written as $f_{\myVec{\theta}}$, which is tuned via
   \begin{align}
       \myVec{\theta}^* &=  \argmin_{\myVec{\theta} \in \Theta}  \mySet{L}_{\Data}(f_{\myVec{\theta}})\notag \\
       &=\argmin_{\myVec{\theta}\in \Theta}\frac{1}{\Ntraining}\sum_{t=1}^{\Ntraining} l(f_{\myVec{\theta}},\Input_t, \Label_t) , 
    \label{eqn:DDOpt}
   \end{align}
   where the last equality is obtained by substituting \eqref{eqn:EmpRisk}.
   In practice, the empirical risk in \eqref{eqn:DDOpt} is often combined with  regularizing terms in order to facilitate training and mitigate overfitting.
   
   \begin{figure*}
    \centering
    \includegraphics[width=\linewidth]{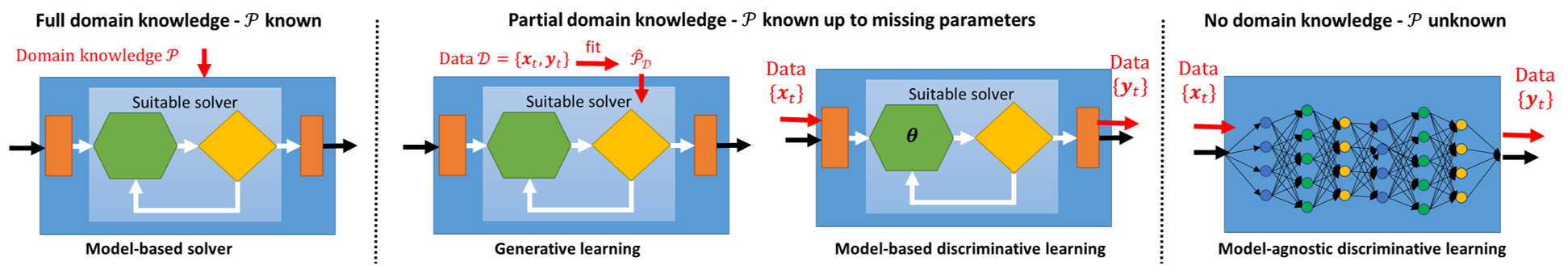}
    \caption{Illustration of different design approaches for inference rules based on domain knowledge and/or data.}
    \label{fig:Overall1}
\end{figure*}

   One can further divide discriminative models into the following categories:
   \begin{enumerate}
       \item {\em Model-based discriminative models}, where one has some domain knowledge that indicates what structure should the inference rule take. This allows using inference mappings with a relatively small number of parameters that are specific to the problem at hand. For instance, in the example presented in the sequel, we consider estimation in a jointly Gaussian setting, for which a suitable inference rule is known to take a linear form.
       \item {\em Model-agnostic discriminative models}, which use highly-parameterized inference rules that can realize a broad set of abstract mappings. This is the common practice in deep learning for inference problems, which infer using \acp{dnn} trained from massive data sets.  
   \end{enumerate}
\end{enumerate}

The different design approaches are illustrated in Fig.~\ref{fig:Overall1}. 
\textcolor{NewColor}{Many tasks encountered in practical applications can in fact be tackled using any of these approaches. For instance, super-resolution, which is a fundamental task in biomedical imaging and optics, can be tackled in a model-based manner by treating it as a fully-known sparse recovery problem, while data can be leveraged to estimate the parameters of the sparse recovery setup via {generative learning}; alternatively, one can use data to directly learn the super-resolution solver using deep unfolding \cite{monga2021algorithm} as a form of model-agnostic discriminative learning, or by training a model-agnostic discriminative model, e.g., a \ac{dnn}. A  detailed account of these examples (and more) can be found in \cite{shlezinger2022model}.}


\subsection{Inference with Partially-Known Generative Distributions}
The unprecedented success of deep learning in areas such as computer vision and natural language processing \cite{lecun2015deep} notably boosted the popularity of model-agnostic discriminative models that rely on abstract, purely data-driven pipelines, trained with massive data sets, for inference. Specifically, by letting $f_{\myVec{\theta}}$ be a \ac{dnn} with parameters $\myVec{\theta}$, one can train inference rules from data via \eqref{eqn:DDOpt} that operate in scenarios where analytical models are unknown or highly complex \cite{Bengio09learning}. For instance, a statistical model $\Distribution$ relating an image of a dog and the breed of the dog is likely to be intractable, and thus, inference rules which rely on full domain knowledge or on estimating $\Distribution$ are likely to be inaccurate. However, the abstractness and extreme parameterization of \acp{dnn} results in them  often being  treated as black boxes, while the training procedure of \acp{dnn} is typically lengthy, computationally intensive, and requires  massive volumes of data. Furthermore, understanding how their predictions are obtained and how reliable they are tends to be quite challenging, and thus, deep learning lacks the interpretability, flexibility, versatility, and reliability of model-based techniques.

Unlike conventional deep learning domains, such as computer vision and natural language processing, in signal processing, one often has access to some level of reliable domain knowledge. Many problems in signal processing applications are characterized by faithful modeling based on the understanding of the underlying physics, the operation of the system, and models backed by extensive measurements. Nonetheless, existing modeling often includes parameters, e.g., channel coefficients and noise energy, which are specific to a given scenario, and are typically unknown in advance, though they can be estimated from data. 
The key question in scenarios involving such partial domain knowledge in addition to training data is which of the following design approaches is preferable:
\begin{enumerate}
    \item The {\em generative learning approach}, which seeks the parameter vector 
    that best matches the data, and then sets the inference rule to minimize the risk with the estimated distribution as in~\eqref{eqn:MBOpt}.
    \item The {\em discriminative learning approach}, which formulates the inference rule for a given parameter vector, and then uses $\Data$ to directly optimize the resulting mapping as in~\eqref{eqn:DDOpt}. 
\end{enumerate}
We tackle this fundamental question using a simple  tractable scenario of linear estimation with partial domain knowledge, where it is known that the setting is linear, yet the exact linear mapping is unknown. In this setting, which is mathematically formulated in the following section, both approaches can be explicitly derived and analytically compared.

\subsection{Case Study: Linear Estimation with Partially-Known Measurement Models}
\label{example_linear}
To provide an analytically tractable comparison between the aforementioned approaches to jointly leverage domain knowledge and data in inference, we consider a linear estimation scenario. Such scenarios are not only simple and enable rigorous analysis, but also correspond to a broad range of statistical estimation problems that are commonly encountered in signal processing applications. Here, the input $\Input$ and the target $\Label$ are real-valued vectors  taking values in $\InputSpace = \mathbb{R}^{N_x}$ and in $\LabelSpace = \mathbb{R}^{N_y}$, respectively. 
The loss measure is the squared-error loss, i.e., $ l_{\rm MSE}(f,\Input,\Label) = \|\Label - f(\Input)\|_2^2$.

\begin{figure}
    \centering
    \includegraphics[width=\columnwidth]{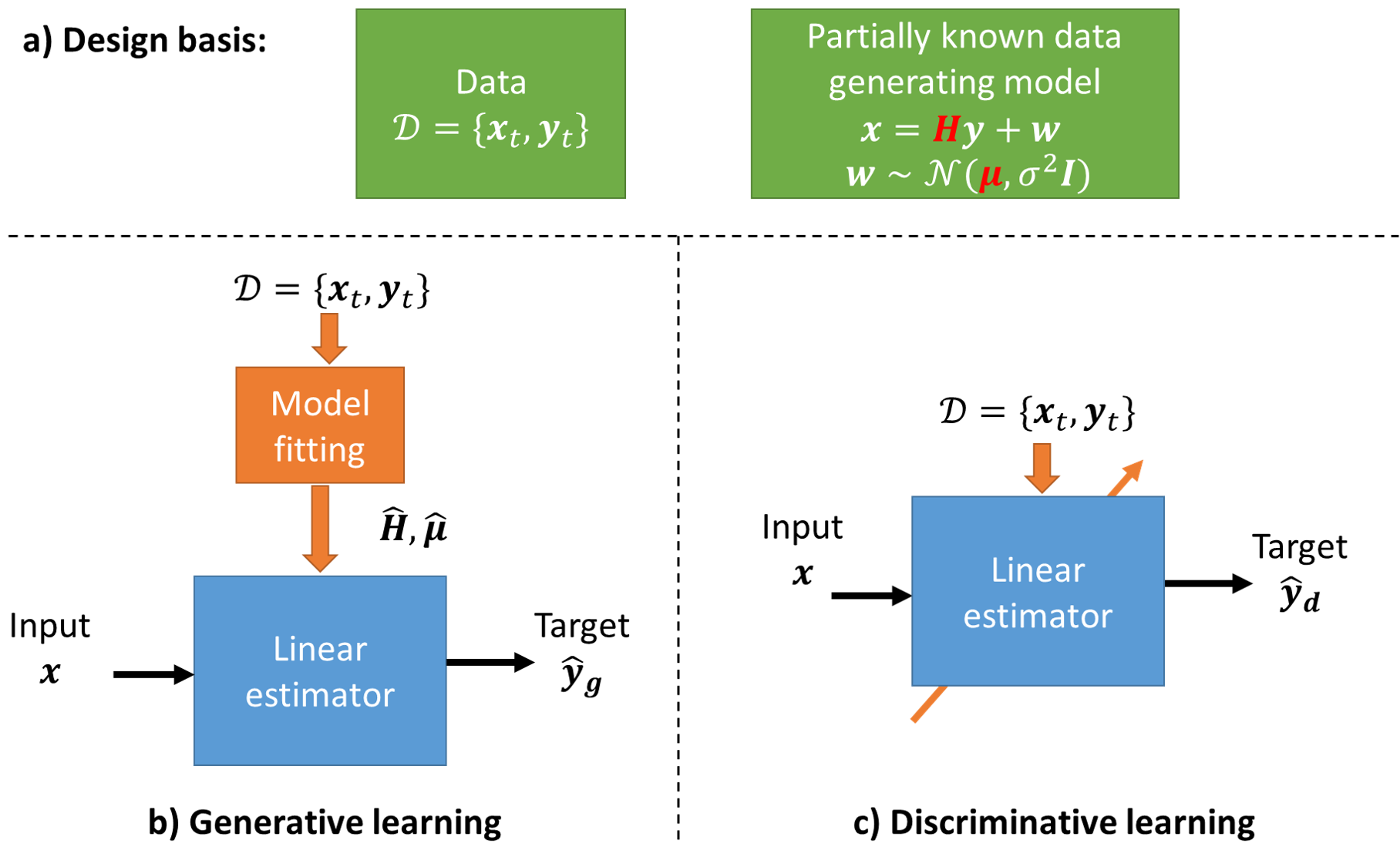}
    \caption{Illustration of the linear estimation setting: $a)$ available data and domain knowledge, where \textcolor{NewColor}{the unknown model parameters are highlighted in red fonts};  $b)$ generative learning based estimator; and $c)$ discriminative learning based estimator.}
    \label{fig:Linear1}
\end{figure}

In the considered setting, one has prior knowledge that the target $\Label$ admits a Gaussian distribution with mean $\myVec{\mu}_y$ and covariance $\myMat{C}_{\Label\Label}$, i.e. 
$\myVec{y}\sim\mathcal{N}(\myVec{\mu}_y,\myMat{C}_{\Label\Label})$,
and that the measurements follow a linear model 
\begin{equation}
\label{eqn:IORel}
 \Input = \myMat{H}\Label + \myVec{w}, \quad \myVec{w}\sim\mathcal{N}(\myVec{\mu},\sigma^2\myMat{I}), 
\end{equation}
i.e., $\myVec{w}$ is a Gaussian noise with mean $\myVec{\mu}$ and covariance matrix $\sigma^2\myMat{I}$, and is assumed to be  independent of  $\Label$. 
However, the available domain knowledge is partial in the sense that  the parameters $\myMat{H}$ and $\myVec{\mu}$ are unknown.
Consequently, while the generative distribution $\Distribution$ is known to be jointly Gaussian, its parameters are unknown. 
Thus, conventional Bayesian estimators, such as the minimum MSE (MMSE) and LMMSE estimators, cannot be implemented for this case.
Nonetheless, we are given access to a data set $\Data =\{\Input_t, \Label_t\}_{t=1}^{\Ntraining}$ comprised of i.i.d. samples drawn from $\Distribution$. An illustration of the  setting is depicted in Fig.~\ref{fig:Linear1}(a). The goal is to utilize the available domain knowledge and the data $\Data$ in order to formulate an inference mapping that achieves the lowest generalization error, i.e., minimizes \eqref{eqn:risk} with the squared-error loss. 

\section{Solution: Data-Aided Linear Estimators}
In the following we exemplify a generative learning estimator (in Subsection \ref{Gen_solution_subsection})  and a discriminative learning estimator (in Subsection \ref{Dis_sub}), which both aim to recover the random signal $\Label$ from  the observed $ \Input$ for the partially known jointly-Gaussian setting  described in Subsection \ref{example_linear}.
To this end, we use the following notations of sample means  
\be
\label{sample_means}
\bar{\Input}\define \frac{1}{\Ntraining}\sum_{t=1}^{\Ntraining} \Input_t,~~~ \bar{\Label}\define\frac{1}{\Ntraining}\sum_{t=1}^{\Ntraining} \Label_t,
\ee
and the sample covariance/cross-covariance matrices 
\begin{subequations}
\label{eqn:sampCov}
\begin{equation} \label{hat_C_theta_x}
	 \hat{\myMat{C}}_{\Label\Input} = \frac{1}{\Ntraining} \sum_{t=1}^{\Ntraining} (\Label_t - \bar{\Label})(\Input_t -\bar{\Input})^T,
\end{equation}
\begin{equation} \label{hat_C_x}
	 \hat{\myMat{C}}_{\Label\Label} =\frac{1}{\Ntraining} \sum_{t=1}^{\Ntraining} (\Label_t - \bar{\Label})(\Label_t - \bar{\Label})^T,
\end{equation}
and
\begin{equation} \label{hat_C_y}
	 \hat{\myMat{C}}_{\Input\Input} = \frac{1}{\Ntraining} \sum_{t=1}^{\Ntraining} (\Input_t - \bar{\Input})(\Input_t - \bar{\Input})^T.
\end{equation}
\end{subequations}

\subsection{Generative Learning Estimator}
\label{Gen_solution_subsection}
The generative approach uses data to estimate the missing domain knowledge parameters. 
In order to estimate the distribution based on the model in \eqref{eqn:IORel}, we need to estimate the matrix $\myMat{H}$ and  the noise mean $\myVec{\mu}$ from the training data, and then use the estimates denoted $\hat{\myMat{H}}$ and $\hat{\myVec{\mu}}$ to form the linear estimator, as illustrated in Fig.~\ref{fig:Linear1}(b). 

The unknown $\myMat{H}$ and $\myVec{\mu}$ are fitted to the data using the maximum likelihood rule. Letting $\Distribution(\myVec{x},\myVec{y};\myMat{H}, \myMat{\mu})$ be the joint distribution of $\myVec{x}$ and $\myVec{y}$ for given $\myMat{H}$ and $\myVec{\mu}$, the log-likelihood can be written as
\begin{eqnarray}
\label{eqn:loglike1}
 \log \Distribution(\Input_t,\Label_t;\myMat{H},\myVec{\mu})=
  \log \Distribution(\Input_t|\Label_t;\myMat{H},\myVec{\mu})+ \log \Distribution(\Label_t)\nonumber\\
  ={\text{const}}
 -\frac{1}{\sigma^2} \|\Input_t-\myMat{H}\Label_t -\myVec{\mu}\|_2^2.
\end{eqnarray}
In \eqref{eqn:loglike1}, ${\text{const}}$ denotes a constant term, which is not a function of the unknown parameters $\myMat{H}$ and $\myVec{\mu}$. 

Under the assumption that $\Data =\{\Input_t, \Label_t\}_{t=1}^{\Ntraining}$ is comprised of  i.i.d. samples drawn from a Gaussian generative distribution, $\Distribution$, the maximum likelihood estimates are given by
\begin{equation}
\label{eqn:loglike}
    \hat{\myMat{H}},\hat{\myVec{\mu}} = \argmax_{\myMat{H}\in {\mathbb{R}}^{N_x\times N_y}, \myVec{\mu}\in {\mathbb{R}}^{N_x}} \sum_{t=1}^{\Ntraining}\|\Input_t-\myMat{H}\Label_t -\myVec{\mu}\|_2^2.
\end{equation}
The solutions to \eqref{eqn:loglike} are given by  \textcolor{NewColor}{\cite[Section 3.3]{theodoridis2020machine}}
\begin{equation}
\label{eqn:hat_H}
\hat{\myVec{\mu}} = \bar{\Input}-
 \hat{\myMat{H}}\bar{\Label}, \quad 
\hat{\myMat{H}} = \hat{\myMat{C}}_{\Input\Label} \hat{\myMat{C}}_{\Label\Label}^{-1},
\end{equation}
where 
$ \hat{\myMat{C}}_{\Input\Label}= \hat{\myMat{C}}_{\Label\Input}^T$, which is defined in
\eqref{hat_C_theta_x}, and it is assumed that $ \hat{\myMat{C}}_{\Label\Label}$ from \eqref{hat_C_x} is a non-singular matrix. 
By substituting the estimators from \eqref{eqn:hat_H} in \eqref{eqn:IORel}, the estimated $\hat{\Distribution}_{\Data}$ is obtained from the linear model 
\begin{equation}
\label{eqn:IORel_est}
 \Input = \hat{\myMat{H}}\Label + \tilde{\myVec{w}}, \quad \tilde{\myVec{w}}\sim\mathcal{N}(\hat{\myVec{\mu}},\sigma^2\myMat{I}).
\end{equation}

Having estimated the generative model, we proceed to finding the inference rule which minimizes the  risk function with respect to $\hat{\Distribution}_{\Data}$, i.e.,
    \begin{equation}
        \label{eqn:MBOpt_linear}
        f^* = \argmin_{f\in \mySet{F}} 
        \E_{(\Input,\Label)\sim \hat{\Distribution}_{\Data}} \{\|\Label - f(\Input)\|_2^2\},
    \end{equation}
    where we substitute the  squared-error loss 
    and  \eqref{eqn:risk} into \eqref{eqn:MBOpt}.
    Since the estimated distribution,  $\hat{\Distribution}_{\Data}$,
is a jointly Gaussian distribution,
the solution of  \eqref{eqn:MBOpt_linear} is the LMMSE estimator under the estimated distribution, which is given by 
  \begin{align}
        &\hat{\myVec{y}}_g=f^*( \Input) =
        \E_{(\Input,\Label)\sim \hat{\Distribution}_{\Data}} \{\Label | \Input\} \notag \\ 
        &\stackrel{(a)}{=} {\myVec{\mu}}_y \! + \!{\myMat{C}}_{\Label \Label}\hat{\myMat{H}}^T ( \hat{\myMat{H}} {\myMat{C}}_{\Label \Label}\hat{\myMat{H}}^T \! + \!\sigma^2\myMat{I})^{-1} ( \Input \! - \!\bar{\Input}\! - \!\hat{\myMat{H}}({\myVec{\mu}}_y\! - \!\bar{\Label})) \notag \\
        &\stackrel{(b)}{=}
    {\myVec{\mu}}_y \! + \!( \hat{\myMat{H}}^T\hat{\myMat{H}}\! + \!\sigma^2 {\myMat{C}}_{\Label \Label}^{-1})^{-1} \hat{\myMat{H}}^T( \Input \! - \!\bar{\Input}\! - \!\hat{\myMat{H}}({\myVec{\mu}}_y\! - \!\bar{\Label})).
    \label{eqn:MBOpt_linear2_5}
  \end{align}
  Here, $(a)$ follows from the estimated jointly Gaussian model \eqref{eqn:IORel_est}, and  $(b)$ is obtained using the  matrix inversion lemma, as long as $ \hat{\myMat{H}}^T\hat{\myMat{H}}\! + \!\sigma^2 {\myMat{C}}_{\Label \Label}^{-1}$ is invertible.
  
     Applying the matrix inversion lemma  requires the computation of the inverse of an $N_y \times N_y$ matrix instead of $N_x \times N_x$ matrix as in the direct expression for the LMMSE estimate. This contributes to the computational complexity when $N_y < N_x$, i.e., the target signal is of a lower dimensionality compared with the input signal. By substituting \eqref{eqn:hat_H} in \eqref{eqn:MBOpt_linear2_5}, one obtains the generative learning estimator as \textcolor{NewColor}{in \eqref{eqn:MBOpt_linear3}.}  {\textcolor{NewColorR2}{It can be seen in \eqref{eqn:MBOpt_linear3} that in the general case, the generative learning estimator is a function of both 
  the empirical covariance of $\Label$, $\hat{\myMat{C}}_{\Label\Label}$, and its true covariance, ${\myMat{C}}_{\Label\Label}$.}}
     \begin{figure*}
     \color{NewColor}
          \beqna
        \label{eqn:MBOpt_linear3}
       \hat{\myVec{y}}_g=f^*( \Input) 
        = {\myVec{\mu}}_y +(\hat{\myMat{C}}_{\Label\Label}^{-1}
        \hat{\myMat{C}}_{\Label\Input} 
        \hat{\myMat{C}}_{\Input\Label}\hat{\myMat{C}}_{\Label\Label}^{-1}+ \sigma^2{\myMat{C}}_{\Label \Label}^{-1})^{-1}  \hat{\myMat{C}}_{\Label\Label}^{-1}
        \hat{\myMat{C}}_{\Label\Input} ( \Input -\bar{\Input}-\hat{\myMat{C}}_{\Input\Label} \hat{\myMat{C}}_{\Label\Label}^{-1}({\myVec{\mu}}_y-\bar{\Label})).
    \eeqna
    \hrule
    \color{black}
     \end{figure*}

\subsection{Discriminative Learning Estimator}
\label{Dis_sub}
In this subsection, we consider a discriminative learning approach for the considered estimation problem. Here, the partial domain knowledge regarding the underlying joint Gaussianity indicates that the estimator should take a linear form, i.e., 
\be
\label{f_linear}
f_{\myVec{\theta}}(\Input)=\myMat{\Amat}\Input+\myVec{b},
\ee
where $\myVec{\theta}=\{\myMat{\Amat},\myVec{b}\}$. For the considered parametric model, the available data is used to directly identify the parameters which minimize the empirical risk, as illustrated in Fig.~\ref{fig:Linear1}(c), i.e., 
\begin{align} 
       \myVec{\theta}^* 
       &=  \argmin_{\myVec{\theta}\in \Theta}  \frac{1}{\Ntraining}\sum_{t=1}^{\Ntraining} 
       \|\Label_t - f_{\myVec{\theta}}(\Input_t)\|_2^2 \notag \\
        &=  \argmin_{\myMat{\Amat}\in {\mathbb{R}}^{N_y\times N_x},\myVec{b}\in{\mathbb{R}}^{N_y}} \frac{1}{\Ntraining}\sum_{t=1}^{\Ntraining} 
       \|\Label_t - \myMat{\Amat}\Input_t-\myVec{b}\|_2^2.
         \label{eqn:DDOpt_linear2}
   \end{align}

Since \eqref{eqn:DDOpt_linear2} is a convex function of $\myMat{\Amat}$ and $\myVec{b}$, the optimal solution is obtained by equating the derivatives of \eqref{eqn:DDOpt_linear2} w.r.t. $\myMat{\Amat}$ and $\myVec{b}$ to zero, which results in
\begin{equation} \label{opt_A}
	\myVec{b}^* = \bar{\Label}  -  \myMat{\Amat}^*\bar{\Input},
 \quad 
    \myMat{\Amat}^* = \hat{\myMat{C}}_{\Label\Input} \hat{\myMat{C}}_{\Input\Input}^{-1},
\end{equation}
where the sample means and sample covariance matrices are defined in 
\eqref{sample_means} and \eqref{eqn:sampCov}.
{\textcolor{NewColor}{It is noted that the solution in \eqref{opt_A} is valid and unique if and only if $ \hat{\myMat{C}}_{\Input\Input}$ is a non-singular matrix (see Sections 3.3 and 4.2 in \cite{theodoridis2020machine}). This  generally holds for $\sigma^2>0$ when $\Ntraining>N_{x}$. }}
By substituting  \eqref{opt_A}
into \eqref{f_linear}, we obtain the discriminative learned estimator, which is given by 
\begin{equation} \label{Monte_Carlo_LMMSE}
\hat{\Label}_d= f_{\myVec{\theta}^*}(\Input)=
 \hat{\myMat{C}}_{\Label\Input} \hat{\myMat{C}}_{\Input\Input}^{-1}(\Input-\bar{\Input})+\bar{\Label}.
\end{equation}
It can be verified that the learned estimator in \eqref{Monte_Carlo_LMMSE} coincides with 
the sample-LMMSE estimator that  is obtained by plugging the sample-mean and sample covariance matrices  into  the LMMSE estimator. 

\section{Discussion and comparison}
By focusing on the simple yet common scenario of linear estimation with a partially-known measurement model, we obtained closed-form expressions for the suitable estimators attained via generative learning and via discriminative learning. This allows us to 
compare the resulting estimators, and thus draw insights into the general approaches of generative versus discriminative learning in the context of signal processing applications. In the following we provide a theoretical analysis of the estimators in Subsection~\ref{ssec:comp_theo}, followed by a qualitative comparison discussed in Subsection~\ref{ssec:comp_Qual}. A stimulative study is presented in Section~\ref{ssec:comp_Exp}.

\subsection{Theoretical Comparison}
\label{ssec:comp_theo}
We next provide an analysis of the estimators obtained via generative learning in \eqref{eqn:MBOpt_linear3} and via discriminative learning in \eqref{Monte_Carlo_LMMSE}, by studying their behavior in asymptotic regimes. To show this, we first study the asymptotic setting where the number of samples $n_t$ is arbitrarily large, after which we inspect the case of high  \ac{snr}, i.e., $\sigma^2\rightarrow0$.

{\bf Asymptotic analysis:}
Let us inspect the derived estimators in the asymptotic case when  $ \Ntraining\rightarrow \infty$ and the data $\Data$ is comprised of i.i.d. samples (i.e., ergodic and stationarity  scenario). 
In this case, the discriminative learning estimator
$\hat{\Label}_d$ in 
\eqref{Monte_Carlo_LMMSE} converges to the LMMSE estimator 
i.e.,
\begin{equation} \label{LMMSE}
\hat{\Label}_d= 
 {\myMat{C}}_{\Label\Input} {\myMat{C}}_{\Input\Input}^{-1}(\Input-\myVec{\mu}_x)+\myVec{\mu}_y,
\end{equation}
where $ {\myMat{C}}_{\Label\Input},{\myMat{C}}_{\Input\Input}$ are the true  covariance matrices, and $\myVec{\mu}_x,\myVec{\mu}_y$ are the true expected values.

Similarly,  for $ \Ntraining\rightarrow \infty$
  the generative learning estimator $ \hat{\Label}_g $ stated in \eqref{eqn:MBOpt_linear3}  converges to 
\begin{align}
\hat{\Label}_g = 
       ({\myMat{C}}_{\Label\Label}^{-1}
        {\myMat{C}}_{\Label\Input} 
        {\myMat{C}}_{\Input\Label}{\myMat{C}}_{\Label\Label}^{-1}+& \sigma^2{\myMat{C}}_{\Label \Label}^{-1})^{-1} {\myMat{C}}_{\Label\Label}^{-1}
        {\myMat{C}}_{\Label\Input} 
        \nonumber\\ \times( \Input -\myVec{\mu}_x)+ {\myVec{\mu}}_y.
     \label{eqn:MBOpt_linear4}  
\end{align}
    When the linear model in \eqref{eqn:IORel} holds, it can be verified that 
    \be
    \label{asy1}
    {\myMat{C}}_{\Input\Label}=\myMat{H}{\myMat{C}}_{\Label\Label}, \quad  {\myMat{C}}_{\Input\Input}=\myMat{H}{\myMat{C}}_{\Label\Label}\myMat{H}^T+\sigma^2\myMat{I}.
    \ee
By substituting \eqref{asy1} into \eqref{eqn:MBOpt_linear4} and using  the matrix inversion lemma again, we obtain 
\begin{align}
\hat{\Label}_g
&=(\myMat{H}^T
        \myMat{H}+ \sigma^2{\myMat{C}}_{\Label \Label}^{-1})^{-1} 
        \myMat{H}^T( \Input- \myVec{\mu}_x)+ {\myVec{\mu}}_y
\nonumber\\
 &={\myMat{C}}_{\Label \Label} \myMat{H}^T(\myMat{H}{\myMat{C}}_{\Label \Label}
        \myMat{H}^T+ \sigma^2\Imat)^{-1} 
        ( \Input -\myVec{\mu}_x)+ {\myVec{\mu}}_y
   \nonumber\\
   & =   {\myMat{C}}_{\Label  \Input} {\myMat{C}}_{\Input\Input}^{-1}
  ( \Input -\myVec{\mu}_x) +\myVec{\mu}_y.
   \label{eqn:MBOpt_linear5}
\end{align}
Thus, asymptotically, if the true generative model is linear and given by \eqref{eqn:IORel}, then the two estimators coincide.

{\bf Asymptotic analysis under misspecified (nonlinear) model:}
Often in practice, 
the generative model is nonlinear. 
For instance, consider the case where 
the true generative model is not the linear one in \eqref{eqn:IORel}, but is given by  
\begin{equation} \label{Model}
	\myVec{x} = \myVec{g}(\myMat{H},\myVec{y}) + \myVec{w},
\end{equation}
where the measurement function, $\gvec: {\mathbb{R}}^{N_y} \rightarrow {\mathbb{R}}^{N_x}$ is a nonlinear function, 
and the statistical properties of $\myVec{y}$ and $\myVec{w}$ are the same as described in Subsection \ref{example_linear}.
Although the model is nonlinear, the estimator may be designed based on the linear model in \eqref{eqn:IORel}, either due to mismatches, or due to intentional linearization carried out  to simplify the problem.
 
When
 the true generative model is nonlinear (i.e. under misspecified model), then the two linear estimators $\hat{\Label}_g$ and $\hat{\Label}_d$ are different even asymptotically. 
The asymptotic estimators in this case are given by \eqref{LMMSE} and \eqref{eqn:MBOpt_linear4}, but  \eqref{asy1} does not hold, and thus, $\hat{\Label}_g$ here does not coincide with the LMMSE estimator, as it does in \eqref{eqn:MBOpt_linear5} for the linear generative model.
In this case, the discriminative model approach is asymptotically preferred being the LMMSE estimator, while the generative learning approach, which is based on a mismatched generative model, yields a  linear estimate that differs from the LMMSE estimator, and is thus sub-optimal. 

{\bf High \ac{snr} regime:}
Another setting in which 
one can rigorously compare the estimators
is in the high \ac{snr} regime, where 
$\sigma^2\rightarrow 0$. 
In the following analysis, we focus on settings where $N_y \geq N_x$, i.e., the number of input variables is at most the number of target variables. 
\textcolor{NewColorR2}{Specifically, when $N_y > N_x$ the input covariance becomes singular when $\sigma^2\rightarrow 0$, and the corresponding estimators, as well as the MMSE estimate, may diverge.}
%
In this case, the data satisfies the model from \eqref{eqn:IORel}, i.e., $\Input_t \xrightarrow[\sigma^2 \to 0]{}  \myMat{H}\Label_t + \myVec{\mu}$. Thus, the sample mean in \eqref{sample_means} satisfies
\be
\label{sample_mean_sigma0}
\bar{\Input}\xrightarrow[\sigma^2 \to 0]{} \myMat{H}\bar{\Label}+\myVec{\mu}.
\ee
Similarly, the sample covariance matrices in \eqref{eqn:sampCov} satisfies 
\begin{align} 
	 \hat{\myMat{C}}_{\Label\Input} \xrightarrow[\sigma^2 \to 0]{}  \frac{1}{\Ntraining}& \sum_{t=1}^{\Ntraining} (\Label_t - \bar{\Label})(\myMat{H}\Label_t + \myVec{\mu} - \bar{\Input})^T \notag \\
	 &= \hat{\myMat{C}}_{\Label\Label}\myMat{H}^T.
	 \label{sigma_0_Cyx}
\end{align}
and
\begin{align} 
	 \hat{\myMat{C}}_{\Input\Input} \xrightarrow[\sigma^2 \to 0]{}  \frac{1}{\Ntraining}& \sum_{t=1}^{\Ntraining} (\myMat{H}\Label_t + \myVec{\mu} - \bar{\Input})(\myMat{H}\Label_t + \myVec{\mu} - \bar{\Input})^T 
	\notag \\
	 & =\myMat{H}\hat{\myMat{C}}_{\Label\Label}\myMat{H}^T.
\label{sigma_0_Cxx}
\end{align}

\color{NewColor}
Under the above limit case, we next derive the considered estimators, starting with the generative one. 
{\textcolor{NewColorR2}{First, we note that according to \eqref{eqn:hat_H} and by using  \eqref{sigma_0_Cyx}, one obtains  that under the assumption that $\hat{\myMat{C}}_{\Label\Label}$ is a non-singular matrix \begin{equation}
\label{eqn:hat_H_new}
\hat{\myMat{H}} \xrightarrow[\sigma^2 \to 0]{}  \hat{\myMat{C}}_{\Input\Label} \hat{\myMat{C}}_{\Label\Label}^{-1}=\myMat{H}.
\end{equation}
Thus,
taking the limit $\sigma^2\rightarrow 0$ in the third equality in  \eqref{eqn:MBOpt_linear2_5} and substituting \eqref{eqn:hat_H_new}, we obtain that
  \begin{align}
        \hat{\myVec{y}}_g\xrightarrow[\sigma^2 \to 0]{}  
         {\myVec{\mu}}_y \! + \!{\myMat{C}}_{\Label \Label}{\myMat{H}}^T ( {\myMat{H}} {\myMat{C}}_{\Label \Label}{\myMat{H}}^T \!)^{-1} ( \Input \! - \!\bar{\Input}\! - \!{\myMat{H}}({\myVec{\mu}}_y\! - \!\bar{\Label})).
    \label{eqn:GenEstAsym}
  \end{align}
}}
It is assumed here that $\myMat{H} \textcolor{NewColorR2}{{\myMat{C}}_{\Label \Label}}  \myMat{H}^T $ is non-singular, for which $N_y \geq N_x$ is a necessary condition. Similarly, by substituting \eqref{sigma_0_Cyx} and \eqref{sigma_0_Cxx} in the discriminative learned estimator in  \eqref{Monte_Carlo_LMMSE}, we obtain 
\begin{equation}
\label{y_hat_d_sigma0}
\hat{\Label}_d\xrightarrow[\sigma^2 \to 0]{} 
 \bar{\Label}+\hat{\myMat{C}}_{\Label\Label}\myMat{H}^T (\myMat{H}\hat{\myMat{C}}_{\Label\Label}\myMat{H}^T)^{-1}(\Input-\bar{\Input}).
\end{equation}

\color{NewColor}
While both asymptotic estimators in \eqref{eqn:GenEstAsym} and \eqref{y_hat_d_sigma0} apply the same linear transformation to the input term $( \Input  -  \bar{\Input})$, they vary in how they process the term depending on the target mean {\textcolor{NewColorR2}{and covariance. That is, 
comparing the generative and discriminative cases in \eqref{eqn:GenEstAsym} and \eqref{y_hat_d_sigma0}, we can see that they yield identical expressions except for the use of the true values,  ${\myVec{\mu}}_y$ and ${\myMat{C}}_{\Label \Label}$, in the generative case being replaced by  the empirical values, $\bar{\Label}$ and $\hat{\myMat{C}}_{\Label\Label}$, in the discriminative case. }}
This follows from the prior knowledge of high \ac{snr} available to the generative estimator. While the above estimators differ, they can be shown to coincide when $\bar{\Label}= \myVec{\mu}_y$ {\textcolor{NewColorR2}{and $\hat{\myMat{C}}_{\Label\Label}={\myMat{C}}_{\Label \Label}$}}, which is approached when $n_t$ is sufficiently large, such that the empirical mean $\bar{\Label}$  {\textcolor{NewColorR2}{and empirical covariance $\hat{\myMat{C}}_{\Label\Label}$}} approach the true mean $\myVec{\mu}_y$ {\textcolor{NewColorR2}{and true covariance matrix ${\myMat{C}}_{\Label \Label}$, respectively. }}

\subsection{Qualitative Comparison}
\label{ssec:comp_Qual}

The theoretical comparison in Subsection~\ref{ssec:comp_theo} allows to rigorously identify scenarios in which one approach is preferable over the other, e.g., that discriminative learning is more suitable for handling modeling mismatches. 
Another aspect in which the estimators are comparable is in their sample complexity. 
The discriminative linear estimator in \eqref{Monte_Carlo_LMMSE} requires the computation of the inverse sample covariance matrix of $\Input$, $ \hat{\myMat{C}}_{\Input\Input}$, from \eqref{hat_C_y}.  On the other hand, the generative linear estimator in \eqref{eqn:MBOpt_linear3}
 requires the computation of the inverse sample covariance matrix of $\Label$, $ \hat{\myMat{C}}_{\Label\Label}$, in \eqref{hat_C_x}. 
The dimensions of the matrices $ \hat{\myMat{C}}_{\Input\Input}$ and  $ \hat{\myMat{C}}_{\Label\Label}$ are different. Thus, for a limited dataset, it may be easier to implement one of these estimators and guarantee the stability of the inverse covariance matrix. 

In particular, 
in settings where the sample size ($\Ntraining$) is comparable to the observation dimension ($N_x$), the discriminative sample-LMMSE estimator exhibits severe performance degradation. 
This is because the sample covariance matrix is not well-conditioned in the small sample size regime, and inverting it amplifies the estimation error.
Similarly, if $\Ntraining$ is comparable to the estimated vector dimension ($N_y$), the performance of the generative learning estimator in \eqref{eqn:MBOpt_linear3} degrades. 
In such cases, 
the available information is not enough to reveal a sufficiently good model that fits the data, and it can
be misleading due to the presence of  noise and possible outliers.

A possible additional operational gain of generative learning stems from  the fact that it estimates the underlying distribution, which can be used for other tasks. The discriminative learning approach is highly task-specific, learning only what must be learned to produce its estimate, and is thus not adaptive to a new task.

\section{Numerical Comparison}
\label{ssec:comp_Exp}
In this section we evaluate the considered estimators in comparison with the oracle MMSE, which is the MMSE estimator for the model in \eqref{eqn:IORel} and known $\myMat{H}$\footnote{The numerical study is available online at \url{https://gist.github.com/nirshlezinger1/3e92bc16d28c8f2f7feb5031e32b5618}}. The purpose of this study is to numerically assert the theoretical findings reported in the previous section, and to empirically compare the considered approaches to combine data with partial domain knowledge in a non-asymptotic regime.

We simulate jointly-Gaussian signals via the signal model in \eqref{eqn:IORel} with $N_x=28$ observations and $N_y = 30$ target entries. The target $\Label$ has zero-mean and covariance matrix representing spatial exponential decay, where the $(i,j)$th entry of $\myMat{C}_{\Label\Label}$ is set to $e^{-\frac{|i-j|}{5}}$. The measurement matrix $\myMat{H}$ is generated randomly with i.i.d. zero-mean unit variance Gaussian entries. The results are averaged over $10^4$ Monte Carlo simulations. 

We numerically evaluate the performance of the discriminative learning estimator of \eqref{Monte_Carlo_LMMSE} and the generative learning estimator of \eqref{eqn:MBOpt_linear3}. 
We consider
two scenarios: 1) a setting in which $\myMat{C}_{\Label\Label}$ and $\myVec{\mu}_y$ are accurately known; and 2) a mismatched case in which the estimator approximates $\myMat{C}_{\Label\Label}$ as the identity matrix. Since the discriminative estimator is invariant \textcolor{NewColor}{with respect to} the prior distribution of $\Label$, the presence of mismatch only affects the generative approach. 

\begin{figure}
    \centering
    \includegraphics[width=\columnwidth]{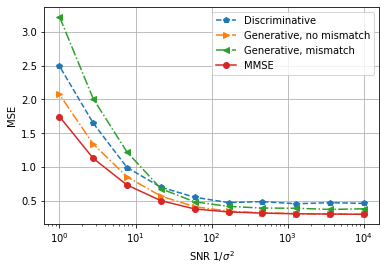}
    \caption{\ac{mse} versus \ac{snr}}
    \label{fig:MSEvsSNR}
\end{figure}

\begin{figure}
    \centering
    \includegraphics[width=\columnwidth]{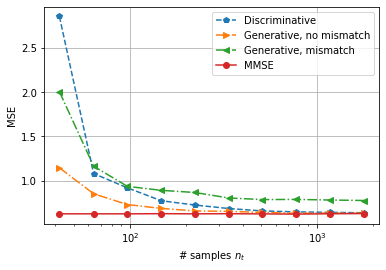}
    \caption{\ac{mse} versus number of samples}
    \label{fig:MSEvsNt}
\end{figure}

The resulting \ac{mse} values versus the \ac{snr} $1/\sigma^2$ for $\Ntraining=100$ data samples are reported in Fig.~\ref{fig:MSEvsSNR}, while Fig.~\ref{fig:MSEvsNt} illustrates the \ac{mse} curves versus $\Ntraining$ for $\sigma = 0.3$. Observing Fig.~\ref{fig:MSEvsSNR}, we note that for the considered setting of small $\Ntraining$, 
the generative estimator, which fully knows $\myMat{C}_{\Label\Label}$, 
outperforms the discriminative approach due to its ability to incorporate the prior knowledge of $\sigma^2$ and of the statistical moments of $\Label$. We also observe that in high \acp{snr}, generative learning allows approaching the MMSE, while  discriminative learning yields a performance gap, which settles with our analysis in the previous section. We note, though, that when repeating this study with $N_x = N_y$, all estimators achieve performance within a minor gap of the MMSE, in the high \ac{snr} regime.  

Nonetheless, in the presence of small mismatches in  $\myMat{C}_{\Label\Label}$, the discriminative approach yields improved MSE, indicating its ability to cope better with modeling mismatches compared with generative learning. In Fig.~\ref{fig:MSEvsNt}, where the \ac{snr} is fixed and finite,  we observe that the effect of a misspecified model does not vanish when the number of samples increases, and the mismatched generative model remains within a notable gap from the MMSE, while both the discriminative learning estimator and the non-mismatched generative one approach the MMSE as $\Ntraining$ grows.  These findings are all in line with the theoretical analysis presented in Subsection~\ref{ssec:comp_theo}.

\section{What we have learned}
\label{sec:Conclusions}
In this lecture note, we reviewed two different approaches to combining partial domain knowledge with data for forming an inference rule. The first approach is related to the machine learning notion of generative learning, which operates in two stages: it first uses data to estimate the missing components in the statistical description of the problem at hand; then, the estimated statistical model is used to form the inference rule. The second approach is task-oriented, leveraging the available domain knowledge to specify the structure of the inference rule, while using data to optimize the resulting mapping in an end-to-end fashion. 

To compare the approaches in a manner that is relevant and insightful to signal processing students and researchers, we focused on a case study representing linear estimation. For such settings, we obtained a closed-form expression for both the generative learning estimator as well as the discriminative learning one. The resulting explicit derivations enabled us to rigorously compare the approaches, and draw insights into their conceptual differences and individual pros and cons.

In particular, we noted that discriminative learning, which uses the available domain knowledge only to determine the inference rule structure, is more robust to mismatches in the mathematical description of the setup. This property indicates the ability of end-to-end learning to better cope with mismatched and complex models. However, when the partial domain knowledge available is indeed accurate, it was shown that generative learning can leverage this prior to \textcolor{NewColor}{operate with few samples and to} improve performance in noisy settings. These findings were not only analytically proven, but also backed by numerical evaluations, which are made publicly available as a highly accessible Python Notebook intended to be used for presenting this lecture in class. 
\textcolor{NewColor}{While our conclusions are rigorously proven for a simple setting of linear estimation in a Gaussian model, they can be empirically shown to generalize to more complex settings. For instance, the robustness of discriminative learning to model mismatches is one of the key motivations for converting model-based algorithms into trainable discriminative models via model-based deep learning~\cite{shlezinger2022model}, while the improved sample complexity of generative learning was also observed in~\cite{ng2001discriminative}.}



\section{Acknowledgments}
The authors thank Arbel Yaniv and Lital Dabush, who helped during the development of the numerical example.

\section{Authors}
\textbf{Nir Shlezinger}  is an assistant professor in the School of Electrical and Computer Engineering at Ben-Gurion University, Israel. He received his B.Sc., M.Sc., and Ph.D. degrees in 2011, 2013, and 2017, respectively, from Ben-Gurion University, Israel, all in electrical and computer engineering. From 2017 to 2019 he was a postdoctoral researcher in the Technion, and from 2019 to 2020 he was a postdoctoral researcher in Weizmann Institute of Science, where he was awarded the FGS prize for outstanding research achievements. His research interests include communications, information theory, signal processing, and machine learning.

\textbf{Tirza Routtenberg} 
(tirzar@bgu.ac.il) is an Associate Professor in the School of Electrical and Computer Engineering at Ben-Gurion University of the Negev, Israel. 
In addition, 
she  has been appointed as the a William R. Kenan, Jr., Visiting Professor for Distinguished Teaching at the Electrical and Computer Engineering Department at
 Princeton University.
She was the recipient of four Best Student Paper Awards at international Conferences. She is currently serving as an Associate Editor of IEEE Transactions on Signal and Information Processing Over Networks and of IEEE Signal Processing Letters. She is a member of the IEEE Signal Processing Theory and Methods Technical Committee. Her research interests include statistical signal processing, estimation and detection theory, graph signal processing, and optimization and signal processing for smart grids.

\bibliographystyle{IEEEtran}
\bibliography{IEEEabrv,refs}

\end{document}

%% file: myshorts.tex
\newcommand{\gvec}{{\bf{g}}}

\newcommand{\Amat}{{\bf{A}}}

\newcommand{\Imat}{{\bf{I}}}

\newcommand{\define}{\stackrel{\triangle}{=}}

\newcommand{\be}{\begin{equation}}
\newcommand{\ee}{\end{equation}}
\newcommand{\beqna}{\begin{eqnarray}}
\newcommand{\eeqna}{\end{eqnarray}}


%% file: Preamble.tex
\definecolor{NewColor}{rgb}{0,0,0}
\definecolor{NewColorR2}{rgb}{0,0,0}

\acrodef{dnn}[DNN]{deep neural network}
\acrodef{map}[MAP]{maximum a-posteriori probability}
\acrodef{snr}[SNR]{signal-to-noise ratio}
\acrodef{mse}[MSE]{mean-squared error}
\acrodef{ista}[ISTA]{iterative soft thresholding algorithm}
\acrodef{lista}[LISTA]{learned \ac{ista}}

\newcommand{\myVec}[1]{{\boldsymbol{#1}}}
\newcommand{\myMat}[1]{{\boldsymbol{#1}}}
\newcommand{\mySet}[1]{\mathcal{#1}}

\newcommand{\Input}{\myVec{x}}
\newcommand{\InputSpace}{\mySet{X}}
\newcommand{\Label}{\myVec{y}}
\newcommand{\LabelSpace}{\mySet{Y}}
\newcommand{\Distribution}{\mySet{P}}
\newcommand{\Data}{\mySet{D}}
\newcommand{\E}{{\mathbb{E}}}
\newcommand{\Ntraining}{n_t}